\documentstyle[amscd,amssymb,verbatim,12pt]{amsart}
\def\dspace{\baselineskip=0.3 in}
\begin{document}
\dspace
\title[ACCELERATION AND DECELERATION ...  ]{ACCELERATION AND DECELERATION
  IN CURVATURE INDUCED PHANTOM MODEL OF THE LATE AND FUTURE UNIVERSE, cosmic COLLAPSE
   AS WELL AS  ITS QUANTUM ESCAPE}

\author{\bf S.K.Srivastava}
{ }
\maketitle
\centerline{ Department of Mathematics,}
 \centerline{ North Eastrn Hill University,}
 \centerline{  Shillong-793022, India}
\centerline{ srivastava@@.nehu.ac.in; sushilsrivastava52@@gmail.com}

\vspace{1cm}

\centerline{\bf Abstract}

Here, cosmology of the late and future universe is obtained from $f(R)-$
gravity with non-linear curvature terms $R^2$ and $R^3$ ($R$ being the Ricci
scalar curvature). It is different from $f(R)-$dark energy models \cite{snj}
where non-linear curvature terms are taken as gravitational alternative of
dark energy. In the present model, neither linear nor non-linear curvature terms are
taken as dark energy. Rather, dark energy terms are induced by curvature terms
in the Friedmann equation derived from $f(R)-$gravitational equations. It has
advantage over $f(R)-$dark energy models in the sense that the present model
satisfies WMAP results and expands as $\sim t^{2/3}$ during matter
dominance. So, it does not have problems due to which $f(R)-$dark energy
models are criticised in \cite{lds}. Curvature-induced dark energy, obtained
here,  mimics phantom. Different phases of this model, including acceleration
and deceleration during phantom phase, are investigated
here. It is found that expansion of the universe will stop at the age $(3.87
t_0 + 694.4) {\rm kyr}$ ($t_0$ being the present age of the universe) and, after
this epoch, it will contract and collapse by the time $(336.87 t_0 +
694.4) {\rm kyr}$. Further, it is shown that universe will escape predicted
collapse (obtained using classical mechanics) on making quantum gravity
corrections relevant near collapse time due to extremely high energy density and
large curvature analogous to the state of very early universe. Interestingly,
cosmological constant is also induced here, which is extremely small in
classical domain, but very high in the quantum domain.


\vspace{1cm}

\centerline {\underline{\bf 1. Introduction}}
 \smallskip

Astrophysical observations, made at the turn of the last century \cite{sp, ag}
show conclusive evidence for acceleration in
the late universe, which is still a challenge for cosmologists. Theoretically,
it is found that dark energy (DE) violating 
{\em strong energy condition} (SEC) or {\em weak energy condition} (WEC) is
responsible for it. So, in the recent past, many DE models violating one of
these conditions, were proposed to
explain the {\em late cosmic acceleration} taking some exotic matter being the
possible DE source.   A comprehensive review of
these 
models is available in \cite{ejc}. As the exotic matter is not known, 
different scalar fields with special attention to quintessence,
phantom, K-essence and tachyons, were tried upon to probe actual nature of the
matter resposible for required DE \cite{ejc}. Apart from, field-theoretic models, some
fluid dynamical models were also proposed. Chaplygin gas and
generalized Chaplygin gas models of DE are prominant among these, as it has
super-symmetric generalization and negative pressure needed for violation of SEC or
WEC\cite{ejc, sksch}. 

As no satisfactory model came up, it was thought that curvature, being the
source of gravitation, could also be a source of DE. The Einstien-Hilbert term
$R/16 \pi G$ ($R$ being the scalar curvature and $G$ being the gravitational
constant) yields the geometrical component in Einstein's gravitational equation, so
it was realized to use non-linear terms of curvature replacing  matter lagrangian in the Einstein's gravitational action. Motivated by this idea,
S. Capozziello {\em et al.} and S.M. Carroll {\em et al.} proposed
non-linear terms of curvature $R^{-n}$ with $n >0$ as a possible
source of DE \cite{cap}. Although this model explained late cosmic
acceleration, it exhibited instability and failed to satisfy solar system
constraints. Further, this idea was taken up by Nojiri and Odintsov and models
for gravitaional alternative of DE were proposed taking different forms of
$f(R)$ other than $R^{-n}$. These improved models satisfied solar system
constraints exhibiting late 
cosmic acceleration for small curvature and early inflation for large
curvature. Thus, in $f(R)-$ {\em dark energy models},  non-linear curvature
terms were considered as an alternative for DE \cite[for detailed
review]{snj}.  Recently, Amendola {\em et al.} have criticized $f(R)-$ {\em
  dark energy models} and have shown that these models with dominating powers of $R$
 for large or small $R$ can not yield viable cosmology as results contradict the
 standard model and do not satisfy Wilkinson Microwave Anisotropy
 Probe (WMAP) observations, though these
 models pass solar system constraints and explain late acceleration \cite{lds}. Moreover,
 it is also shown that
 the most popular model with $f(R) = R + \alpha R^m + \beta R^{-n} (m >0, n
 >0)$ considered in \cite{snj}
 is unable to  produce matter in the late universe prior to the beginning of
 late acceleration  \cite{lds}.

The recent developements against $f(R)-$ {\em dark energy models} indicate the
need to think for some alternative approach to get DE from curvature. Such an
approach has been proposed in \cite{sks07a, sks06, sks07b, sks08a, sks08b} and
it is adapted here too. This
apporach yields a cosmology emerging from $f(R)-$ {\em gravity} satisfying
WMAP results as well as results of the standard cosmological model. On the
contrary, these results (WMAP results as well as results of the standard
cosmological model) are not staisfied by $f(R)-$ {\em dark energy models}
well as it is noted in \cite{lds}. There is a crucial difference between the  cosmology from $f(R)-$ {\em gravity}, being addressed here as well
as in \cite{sks07a, sks06, sks07b, sks08a, sks08b} and $f(R)-$ {\em dark
  energy models}. In the latter case,  {\em non-linear terms of curvature are
  taken as dark energy lagrangian a priori}. On the contrary, in the former
case, {\em neither linear nor non-linear term is considered as dark
  energy}. Rather, in the present cosmological
model from $f(R)-$ {\em gravity}, DE is induced by linear
(Einstein-Hilbert term) as well as  non-linear terms $R^2$ and $ R^{(2 + r)}$
in the action ( $r$ being a real number such that $(2 + r) >0$). In $f(R)-$ {\em dark energy models},
dark energy terms depend on $f(R)$ terms and its derivative $F = df/dR$. In
the former case, induced DE terms depend on the scale factor $a(t)$ of the
homogeneous and flat of Friedmann - Robertson - Walker (FRW) model of the
universe.

In what follows,  $f(R)-$ {\em gravity} with $R^2$ and $R^3$ is considered. It
is found that curvature terms induce dark energy, dark matter and cosmological
constant in the Friedmann equation (FE) for the late universe, derived from $f(R)-$
gravitational equations. It is interesting to see that curvature-induced dark
energy, obtained here, mimics phantom with equation of state (EOS) 
parameter $ w = - 5/4$. Moreover, FE contains phantom DE term as $\rho_{\rm
  de} [1 - \rho_{\rm de}/2 \lambda]$. The correction term $ - \rho^2_{\rm
  de}/2 \lambda $ with $\lambda$ being the cosmic tension \cite{sks07b,
  sks08b} is analogous to such term in RS-II model based FE \cite{rm} as well as loop
quantum gravity correction \cite{ms}. Like references \cite{sks07b, sks08b},
here also, 
this term is obtained from  $f(R)-$ {\em gravity}. Here, cosmic tension
$\lambda$ is evaluated to be $5.77 \rho^0_{\rm cr}$ with $\rho^0_{\rm cr}$
being the present critical density of the universe. Further, it is shown that
 universe, derived by curvature-induced dark matter, decelerated upto time
 $0.59 t_0$. At this epoch and small red-shift $z =
 0.36$, transition from deceleration to acceleration took
 place. Interestingly, it is noted that as phantom energy density increases,
 effect 
 of  the term $ - \rho^2_{\rm   de}/2 \lambda $ gradually increases. As a
 result, it is found that universe will super-accelerate (expansion with high
 acceleration) during the 
 period $ 0,59 t_0 < t < 2.42 t_0$,  it will accelerate (expansion with low
 acceleration) during the period $2.42 t_0 < t < 3.44 t_0$ 
and, during the period $3.44 t_0 < t < 3.87 t_0$, universe will decelerate even
during the phantom phase. Phantom-dominance will end when $\rho = 2 \lambda =
11.54 \rho^0_{\rm cr}$ at time $t = 3.87 t_0$ causing re-dominance
of dark matter with the re-beginning of decelerated expansion. It is natural
to think that, even during this phase with deceleration (derived by dark
matter), phantom DE will grow with expansion causing $\rho_{\rm
  de} [1 - \rho_{\rm de}/2 \lambda]$ negative due to $ \rho_{\rm de} >2
\lambda$ during re-dominabce of dark matter. In the beginning of this phase,
this term can be ignored, but its effect will grow  causing expansion to stop
at time $t_m = 3.87 t_0 + 694.4 {\rm kyr}$.  As a consequence, universe will retrace back and contract. Further, it
  is found that by the time $i_{\rm col} = 336.87 t_0 + 694.4 {\rm kyr}$,
  universe will 
  collapse and energy density will diverge. It is argued that this state of the
  future universe is analogous to the state of very early universe, where
  energy density is extremely large and curvature is very high. So, like early
  universe, it is reasonable to think that quantum gravity effects will be
  dominant near the collapse 
  time. Motivated by this idea, quantum corrections are made in FE in the
  small interval 
  of time around the time of collapse and it is found that cosmic collapse,
  predicted by classical mechanics, can be avoided. Earlier too, Quantum
  gravity corrections were done in the future phantom universe to avoid finite
  time singularities \cite{eil, nopl, sksgrg}. The paper is organized as
  follows. 

In section 2, phantom-phase of the late and future universe is
investigated. Transition from deceleration to acceleration is probed and
cosmic tension is evaluated. Section 3 contains discussion on
energy conditions during phantom era and its cosequences. In section 4,
re-dominance of dark matter, maximum expansion, contraction and possible
cosmic collapse using the classical approach have been addressed. Avoidance of
collapse, using quantum gravity effects, is demonstrated in sub-section
4(c). Important results are summarized in the last section.

Natural units$(k_B = {\hbar} = c = 1)$ (where $k_B, {\hbar}, c$ have their
usual meaning) are used here. GeV is used as a fundamental unit and we have $1
{\rm GeV}^{-1} = 6.58 \times 10^{-25} sec.$

\bigskip

\centerline {\underline{\bf 2.  Phantom phase of the late universe and future
  universe }}
 
\centerline {\underline{\bf from $f(R)$-gravity}}

\smallskip

Here, the action for $f(R)-$ gravity is taken as 
$$ S = \int {d^4x} \sqrt{-g} \Big[\frac{R}{16 \pi G} + \alpha R^2 + \beta R^3 
 \Big],  \eqno(2.1)$$
where $G = M_P^{-2} (M_P = 10^{19} {\rm GeV}$ is the Planck mass), $\alpha$ is
a dimensionless coupling constant and $\beta$ is a constant having dimension
(mass)$^{-2}$ (as $R$ has mass dimension 2).

Action (1) yields field equations
$$\frac{1}{16 \pi G} (R_{\mu\nu} - \frac{1}{2} g_{\mu\nu} R) + \alpha (2
\triangledown_{\mu} \triangledown_{\nu} R - 2 g_{\mu\nu} {\Box} R -
\frac{1}{2} g_{\mu\nu} R^2 + 2 R R_{\mu\nu} ) $$
$$ + \beta (3 \triangledown_{\mu} \triangledown_{\nu} R^2 - 3 g_{\mu\nu} {\Box}
R^2 - \frac{1}{2} g_{\mu\nu} R^3 + 3 R^2 R_{\mu\nu} )   = 0 \eqno(2.2a)$$ 
using the condition $\delta S/\delta g^{\mu\nu} = 0$. The operator ${\Box}$ is
defined as
$$ {\Box} = \frac{1}{\sqrt{-g}} \frac{\partial}{\partial x^{\mu}}
\Big(\sqrt{-g} g^{\mu\nu} \frac{\partial}{\partial x^{\nu}} \Big). \eqno(2.2b)$$

Taking trace of (2.2a), it is obtained that
$$ - \frac{R}{16 \pi G} - 6 (\alpha + 3 \beta R){\Box} R - 18 \beta
\triangledown^{\mu}R \triangledown_{\mu}R + \beta R^3 = 0 \eqno(2.3)$$

In (2.3), $(\alpha + 3 \beta R)$ emerges as a coefficient of ${\Box} R$ due to
presence of terms $\alpha R^2$ and $\beta R^3$ in the action (2.1). If $\alpha =
0$, effect of $R^2$ vanishes and  effect of $R^3$ is switched
off for $\beta = 0$. So, an {\em effective} scalar curvature ${\tilde R}$ is defined as
$$ \gamma {\tilde R} = \alpha + 3 \beta R , \eqno(2.4)$$
where  $\gamma$ is a constant having dimension (mass)$^{-2}$ being
used for dimensional correction.

Connecting (2.3) and (2.4), it is obtained that
$$ - {\Box}{\tilde R} - \frac{1}{\tilde R}\triangledown^{\mu}{\tilde R}
\triangledown_{\mu} {\tilde R} = \frac{1}{6 \gamma} \Big[\frac{1}{16 \pi G} -
\frac{\alpha^2}{3\beta} \Big] - \frac{1}{6 \gamma} \Big[\frac{1}{16 \pi G} -
\frac{\alpha^2}{9\beta} \Big]\frac{1}{\gamma{\tilde R}}$$
$$ - \frac{\tilde R}{54 \beta} [\gamma {\tilde R} - 3 \alpha ] . \eqno(2.5)$$
 
Experimental evidences \cite{ad} support spatially homogeneous
flat model of the universe 
$$dS^2 = dt^2 - a^2(t) [dx^2 + dy^2 + dz^2] \eqno(2.6)$$
with $a(t)$ being the scale-factor.

In this space-time,  (2.5) is obtained as
$$- {\ddot {\tilde R}} - 3 \frac{\dot a}{a} {\dot {\tilde R}} -\frac{{\dot
{\tilde R}}^2}{{\tilde R}} =  \frac{1}{6 \gamma} \Big[\frac{1}{16 \pi G} -
\frac{\alpha^2}{3\beta} \Big] - \frac{1}{6\gamma} \Big[\frac{1}{16 \pi G} -
\frac{\alpha^2}{9\beta} \Big]\frac{1}{\gamma{\tilde R}}$$
$$ - \frac{\tilde R}{54 \beta} [\gamma {\tilde R} - 3\alpha ] \eqno(2.7)$$
due to spatial homogeneity.

For $a(t)$, being the power-law function of $t$, ${\tilde R} \sim a^{-n}$. For
example,  ${\tilde R} \sim a^{-3}$ for
matter-dominated model. So, there is no harm in taking
$$  {\tilde R} = \frac{A}{a^n} , \eqno(2.8)$$
where $n > 0$ is a real number and $A$ is a constant with mass dimension 2.

Using (2.8) in (2.7) ,it is obtained that
$$\frac{d}{dt} \Big(\frac{\dot a}{a}\Big) + (3 - 2 n) \Big(\frac{\dot a}{a}
\Big)^2 = \frac{ C a^n}{n A } \Big[ 1 - \frac{D a^n}{C} \Big] -
\frac{\gamma A}{54 n \beta} \Big[ \frac{1}{a^n} - \frac{3 \alpha}{\gamma A}
\Big]. $$

In the late universe, $a(t)$ is large, so this equation reduces to 
$$\frac{d}{dt} \Big(\frac{\dot a}{a}\Big) + (3 - 2 n) \Big(\frac{\dot a}{a}
\Big)^2 = \frac{ C a^n}{n A } \Big[ 1 - \frac{D a^n}{C} \Big]  + \frac{
  \alpha}{18 n \beta},  \eqno(2.9a)$$
 where
$$ C = \frac{1}{6 \gamma} \Big[\frac{1}{16 \pi G} - \frac{\alpha^2}{3 \beta}
\Big]    \eqno(2.9b)$$
and
$$ D = \frac{1}{6 \gamma} \Big[\frac{1}{16 \pi G} - \frac{\alpha^2}{9 \beta}
\Big].    \eqno(2.9c)$$

(2.9a) can be re-written as
$${\ddot a} + (2 - 2 n) \frac{{\dot a}^2}{a}
 = \frac{ C a^{n + 1}}{n A } \Big[ 1 - \frac{D a^n}{C} \Big]  - \frac{ A\alpha}{18 n \beta},  \eqno(2.10)$$
which integrates to
$$ \Big(\frac{\dot a}{a} \Big)^2 =  \frac{E}{a^{6 - 4n}} + \frac{2 C}{n A
  } \Big[ \frac{a^n}{(6 - 3n)} - \frac{D a^{2n}}{C (6 - 2n)} \Big] - \frac{3
  \alpha}{9n \beta(6 - 4n)} , \eqno(2.11a)$$
where $E$ is an integration constant having dimension (mass)$^2$.

(2.11a) is the modified Friedmann equation (MFE) giving cosmic dynamics.  Terms, on r.h.s. of this equation,
are imprints of curvature. The first term of these, proportional to $a^{- (6
  -4n)}$ emerge spontaneously. It is interesting to see that this term (the
first term on r.h.s. of (15a)) corresponds to matter density if $n = 3/4$
i.e. for this value of $n$, it reduces to $E a^{-3}$ and yields the density of
non-baryonic matter being spontaneously
induced by curvature. So, we recognize it as dark matter density.

Thus, for $n = 3/4$, (2.11a) looks like 
$$ \Big(\frac{\dot a}{a} \Big)^2 = \Big[\frac{E}{a^3} + \frac{16 \alpha}{729
  \beta} \Big]+ \frac{32 C a^{3/4}}{45 A
  } \Big[ 1 - \frac{5 D a^{3/4}}{6 C} \Big]  .  \eqno(2.11b)$$

On r.h.s. of (11b), there are terms proportional to $a^{3/4}$ and
$a^{3/2}$. If the density term $ \rho_{\rm de} = {4 C a^{3/4}}/{15 A \pi G}$
is put in conservation equation 
$$ {\dot \rho_{\rm de}} + 3 H (\rho_{\rm de} + p_{\rm de}) = 0 \eqno(2.11c)$$
with $p_{\rm de} = w \rho_{\rm de}$, we obtain
$$ w = - \frac{5}{4} < - 1.  \eqno(2.11d)$$
This result shows that $ \rho_{\rm de} = {4 C a^{3/4}}/{15 A \pi G}$ behaves
as phantom dark energy density being induced by $f(R)-$ gravity.

Now, (2.11b) is re-written as
 $$H^2 = \Big(\frac{\dot a}{a} \Big)^2 =  \Big[\frac{8 \pi G}{3} \rho_{\rm dm}
  + 
  \frac{16 \alpha}{729 \beta} \Big] +  \frac{32 C a^{3/4}}{45 A
  } \Big[ 1 - \frac{5 D a^{3/4}}{6 C} \Big]    \eqno(2.11e)$$
with
$$ \rho_{\rm dm} =  \frac{3 E}{8 \pi G a^3}. \eqno(2.11f)$$

Using current value of 
$\rho_{\rm dm}$ as  $0.23 \rho^0_{\rm cr}$ , (2.11f) yields 
$$\rho_{\rm dm} = 0.23 \rho^0_{\rm cr} \Big(\frac{a_0}{a} \Big)^3,
\eqno(2.12a)$$
where $a_0 = a(t_0), {3 E}/{8 \pi G } =  0.23 \rho^0_{\rm cr} a_0^3  $  and  
$$ \rho^0_{\rm cr} = \frac{3 H_0^2}{8 \pi G}  $$
with $H_0 = 100 km/Mpc sec = 2.32 \times 10^{-42} {\rm GeV} h$ being the
current Hubble's rate of expansion and $h = 0.68$. The present age of the universe is
estimated to be $t_0 \simeq 13.7 {\rm Gyr} = 6.6 \times 10^{41} {\rm
  GeV}^{-1}$  \cite{abl}. So,
$$ H_0^{-1} = 0.96 t_0 .  \eqno(2.12b)$$

Further, $a_0$ is normalized as
 $$ a_0 = 1. \eqno(2.12c)$$

Connecting (2.12a) and (2.12c), it is obtained 
$$\rho_{\rm dm} = \frac{0.23 \rho^0_{\rm cr}}{a^3} \eqno(2.12d)$$

WMAP \cite{abl} gives decoupling of matter from radiation at red-shift
$$ Z_d = \frac{1}{a_d} - 1 = 1089. \eqno(2.13)$$
So, it is supposed that dark matter begins to  dominate cosmic dynamics when $a > a_d$.

Now, (2.11e) is re-written as
 $$ \Big(\frac{\dot a}{a} \Big)^2 =  \frac{8 \pi G}{3} \Big[\Big\{\rho_{\rm
  dm} + \frac{2\alpha}{243 \beta \pi G} \Big\} +  \rho_{\rm de} \Big\{ 1 -
  \frac{\rho_{\rm de}}{2 \lambda} \Big\}\Big],    \eqno(2.14a)$$
where
$$ \rho_{\rm de} =  \rho^0_{\rm de} a^{3/4} \eqno(2.14b)$$
with $ \rho^0_{\rm de} = 0.73 \rho^0_{\rm cr} = {4 C }/{15 A \pi G}$ using
$a_0 = 1$ and
$$  \lambda = \frac{3 C \rho^0_{\rm de}}{5 D }        , \eqno(2.14c)$$
where $C$ and $D$ are given by (2.9b) and (2.9c) respectively.

The Friedmann equation (FE) (2.14a) is obtained from $f(R)-$
gravity. Comparing it with general relativity based FE
$$\Big(\frac{\dot a}{a} \Big)^2 =  \frac{8 \pi G}{3} \rho + \frac{\Lambda}{3},
$$
it is obtained that the constant term in (2.14a) behaves as cosmological
constant
$$ \Lambda = \frac{16 \alpha}{243 \beta} \eqno(2.14d)$$
with vacuum energy density
$$ \rho_{\Lambda} = \frac{\Lambda}{8 \pi G} = \frac{2\alpha}{243 \beta \pi G}
. \eqno(2.14e)$$

It is interesting to see that (2.14c) contains a term
$-{(\rho_{\rm de})^2}/{2 \lambda}$ analogous to brane-gravity
correction to the Friedmann equation (FE) for negative brane-tension \cite{rm} and
modifications in FE due to loop-quantum effects \cite{ms}. Here $\lambda$ is
called {\em cosmic tension} \cite{sks06, sks07b}. (2.14c) shows that {\em
  cosmic tension} $ \lambda$ depends on coupling constants $\alpha$ and
$\beta$ in the gravitational action (2.1). Moreover, positive cosmological
constant, too, emerges from curvature.

From (2.12a) and (2.14b), it is found that $\rho_{\rm dm} \sim {a}^{-3}$ and $
\rho_{\rm de} \sim a^{3/4}$. So $\rho_{\rm dm}$ decreases and $\rho_{\rm de}$
increases with expansion of the universe. So, it is natural to think for
values of these to come closer and to be equal at a certain time $t_*$. At
this particular time, we have
$$0.23 {a_*}^{-3} = 0.73 a_*^{3/4} $$
using (2.12a), (2.12c) and (2.14b) as well as $a_* = a(t_*)$.  This equation yields
$$ a_* = \Big(\frac{23}{73}\Big)^{4/15} . \eqno(2.15)$$

It shows that , for $ a < a_* = ({23}/{73})^{4/15} $, $\rho_{\rm dm} >
  \rho_{\rm de} > \rho_{\rm   de}^2$. So,  (2.14a) is approximated as
 $$ \Big(\frac{\dot a}{a} \Big)^2 =  \frac{8 \pi G}{3} \Big[\Big\{\rho_{\rm
  dm} + \frac{2\alpha}{243 \beta \pi G} \Big].     \eqno(2.16)$$

Connecting (2.12a) and (2.16), it is obtained that
$$ \Big(\frac{\dot a}{a} \Big)^2 =  \frac{0.23 H_0^2}{a^3} \Big[1 + \frac{a^3}{B} \Big]     \eqno(2.17a)$$
with 
$$B = \frac{\rho^0_{\rm dm}}{\rho_{\Lambda}}  = \frac{10.48 \beta H_0^2}{\alpha}, \eqno(2.17b)$$
obtained using (2.14e).

(2.17a) integrates to
$$ a = B^{1/2} sinh^{2/3}\Big[ \frac{3 H_0 \sqrt{0.23}}{2 B^{3/2}} (t - t_d) +
sinh^{-1} (a_d/B^{1/3})^{3/2} \Big] \eqno(2.18a)$$
This result is not consistent with the scale factor, obtained in standard
model of cosmology during matter dominance. So, to have a viable cosmology,
it needs to be approximated as
$$ a = a_d \Big[1  + \frac{3 H_0
  \sqrt{0.23}}{2a_d^{3/2}} (t - t_d) \Big]^{2/3}, \eqno(2.18b)$$
which is possible till 
$$\frac{3 H_0 \sqrt{0.23}}{2 B^{3/2}} (t - t_d) +
sinh^{-1} (a_d/B)^{3/2} \lesssim 1 \eqno(2.18c)$$
as $ sinh 1 =1.18 \simeq 1.$ Here $a_d$ is given by (2.13), which is the scale
factor at time $t = t_d = 386 {\rm kyr} = 2.8 \times 10^{-5} t_0.$  Connecting
(2.13),(2.17b), (2.18a) and (2.18c), it is obtained that
$$ \frac{\alpha}{ \beta H_0^2} \gtrsim 10^{-6} . \eqno(2.18d)$$ 

Connecting (2.16e) and $ {\alpha}/{ \beta H_0^2} \simeq 10^{-6}$ from (2.18d),
it is evaluated that
$$ \rho_{\Lambda} = 2.19 \times 10^{-8} \rho^0_{\rm cr} = 5.48 \times 10^{-55}
 {\rm GeV}^4.  \eqno(2.18e)$$

The approximated form (2.17a) is obtained when $ a << a_*$, but, as discussed
above, $\rho_{\rm dm} \simeq \rho_{\rm de}$ when $ a \lesssim a_*.$ So, in the
narrow strip around $ a = a_*$ for $ a < a_*$, (2.14a) is approximated as
 $$ \Big(\frac{\dot a}{a} \Big)^2 =  \frac{8 \pi G}{3} \Big[\Big\{2\rho_{\rm
  dm} + \frac{2\alpha}{243 \beta \pi G} \Big]    \eqno(2.19)$$
using  $\rho_{\rm dm} \simeq \rho_{\rm de} > \rho_{\rm de}^2$ in (2.14a).
 
Connecting (2.12a) and (2.19), it is obtained that
$$ \Big(\frac{\dot a}{a} \Big)^2 =  \frac{0.46 H_0^2}{a^3} \Big[1 +
\Big(\frac{a}{a_*} \Big)^3 \Big]     \eqno(2.20)$$
with $a_*^3 = {\alpha}/{167.67 \beta H_0^2}$. It is because at $ a = a_*$,
mater-dominated phase ends and dark energy dominance begins.

(2.20) integrates to

$$ a^{3/2} =  a_*^{3/2} sinh\Big[ \frac{3 H_0 \sqrt{0.46}}{2 a_*^{3/2}} (t - t_d) +
sinh^{-1} (a_d/a_*)^{3/2} \Big] ,$$
being approximated as
$$ a =  a_d \Big[1  + \frac{3 H_0
  \sqrt{0.46}}{2a_d^{3/2}} (t - t_d) \Big]^{2/3} \eqno(2.21)$$
like the above case.

Using (2.21), it is obtained that
$$a^{3/2}_* \simeq  a_d^{3/2} + \frac{3 H_0 \sqrt{0.46}}{2} (t_* - t_d).$$
This result yields
$$ t_* \simeq 0.59 t_0 \eqno(2.22)$$
using values of $a_*$ (from (2.15)), $H_0^{-1}$ (from (2.12b)) and $t_d$.  

It is discussed above that for $ a > a_*, \rho_{\rm dm} < \rho_{\rm
  de}$. In this case, (2.14a) is approximated as
$$ \Big(\frac{\dot a}{a} \Big)^2 =  \frac{8 \pi G}{3} \rho_{\rm de} \Big\{ 1 -
  \frac{\rho_{\rm de}}{2 \lambda} \Big\}.    \eqno(2.23)$$

Thus, in the late universe, a phantom model is obtained from curvature without
using any source of exotic matter. But this model contains a correction term
$ - 4 \pi G\rho^2_{\rm de}/ 3 \lambda$ due to  curvature-induced cosmic tension $\lambda$
being evaluated below.

Connecting (14b) and (2.23), it is obtained that
$$ H^2 = \Big(\frac{\dot a}{a} \Big)^2 = 0.73 H_0^2 a^{3/2} \Big[ a^{-3/4} -
  \frac{0.73 \rho^0_{\rm cr} }{2 \lambda} \Big\}\Big].    \eqno(2.24)$$

(2.24) integrates to
\begin{eqnarray*}
 a(t) &=&  \Big[ \frac{0.73 \rho^0_{\rm cr}} {2 \lambda} + 
\Big\{\sqrt{ 1.26 - \frac{0.73 \rho^0_{\rm cr}} {2 \lambda}}
\\ &&
- \frac{3}{8} H_0 \sqrt{0.73} (t - t_*)
\Big\}^2 \Big]^{- 4/3}
\end{eqnarray*}
\vspace{-1.7cm}
\begin{flushright} 
(2.25a)
\end{flushright}
as $ a_*^{-3/4} = 1.26$. (25a) shows that phantom model, obtained here, is singularity-free.

From (2.25a), it is obtained that
$$ {\ddot a} = 0.27 H_0^2 a^{5/2} \Big[ \frac{1.7 \rho^0_{\rm cr}}{\lambda} - \frac{11}{3}a^{-3/4} \Big]. \eqno(2.25b)$$
This shows  ${\ddot a} > 0$ , when 
$$\frac{1.7 \rho^0_{\rm cr}}{\lambda} > \frac{11}{3}a^{-3/4}.  \eqno(2.25c)$$

Further, (2.25a) yields
 \begin{eqnarray*}
 1 = a_0 &=&  \Big[ \frac{0.73 \rho^0_{\rm cr}} {2 \lambda^{\rm ph}} + 
\Big\{\sqrt{ 1.26 - \frac{0.73 \rho^0_{\rm cr}} {2 \lambda^{\rm ph}}}
\\ &&
- \frac{3}{8} H_0 \sqrt{0.73} (t_0 - t_*)
\Big\}^2 \Big]^{- 4/3}.
\end{eqnarray*}
\vspace{-1.7cm}
\begin{flushright} 
(2.26)
\end{flushright}

Using (2.22) for $t_*$ in (2.26), $\lambda$ is evaluated as
$$ \lambda = 5.77 \rho^0_{\rm cr} .  \eqno(2.28)$$

(2.25a) exhibits accelerating universe when $t > t_*$. Thus, a transition from
     decleration to acceleration takes place at 
$$t = t_* = 0.59 t_0  \eqno(2.29a)$$
 and red-shift
$$ z_*  = \frac{1}{a_*} - 1 = \Big(\frac{73}{23} \Big)^{4/15} - 1 = 0.36 ,\eqno(2.29b)$$
which is within the
range $0.33 \le Z_* \le 0.59$ given by 16 Type supernova observations \cite{ag}.
(2.25a) shows that universe expands  till $\rho_{\rm de}$ becomes equal to $2
\lambda$ as it grows with
expansion. It happens till $a(t)$ increases to $a_{\rm pe}$ satisfying
$$ {a_{\rm pe}}^{3/4} = \frac{2
  \lambda}{0.73 \rho_{\rm cr}}. \eqno(2.30)$$
Thus, expansion (2.25a) stops at time
$$ t_{\rm pe} = t_*  + \frac{8}{3 H_0 \sqrt{0.73}} \sqrt{1.26 - \frac{0.73 \rho_{\rm cr}}{2 \lambda}}. \eqno(2.31)$$

\bigskip

\centerline{\underline{\bf 3. Cosmic energy conditions as well as acceleration
    and}} 

 \centerline{\underline{\bf deceleration during phantom era}}

\smallskip

In what follows, it is found that correction term, $ - 4 \pi G\rho^2_{\rm de}/ 3 \lambda$ due to
curvature-induced cosmic tension $\lambda$ in (23), change the bahviour of the
phantom model drastically. 

From (2.11c) and (2.23), it is obtained that
$$\frac{\ddot a}{a} = - \frac{4\pi G}{3} \Big[3(\rho + p)\Big[1 -
\frac{\rho}{\lambda}\Big] - 2 \rho\Big\{1 -
\frac{\rho}{2\lambda}\Big\} \Big], \eqno(3.1)$$
 The correction term, in this equation, is caused due to
curvature-induced cosmic tension $\lambda$ in (2.23). This type of equation was
obtained earlier in \cite{nopl, sksgrg} in the context of RS-II model in brane-gravity.

In the GR-based theory, (3.1) looks like
$$\frac{\ddot a}{a} = - \frac{4\pi G}{3} [\rho + 3 P] . \eqno(3.2)$$

Comparing (2.5) and (2.6), the effective pressure density $P$ is given as
$$ \rho + 3 P = 3(\rho + p)\Big[1 -
\frac{\rho}{\lambda}\Big] - 2 \rho\Big[1 -
\frac{\rho}{2\lambda}\Big] . \eqno(3.3)$$
 
Using (2.11d), (3.3) yields the effective pressure density in the
curvature induced phantom model as
$$ P = - \frac{5}{4} \rho  + \frac{7\rho^2}{12\lambda} . \eqno(3.4)$$

(3.4) yields
$$ \rho + P = - \frac{ \rho}{4} + \frac{7\rho^2}{12\lambda}.  \eqno(3.5)$$

This equation shows that WEC is violated when ${\rho} < 3/7 {\lambda} = 2.47 \rho^0_{\rm
  cr}$ with $\lambda$ given by (2.28). Moreover , $\rho + P = 0$ for ${\rho}  =
2.47 \rho^0_{\rm   cr}$ and $\rho + P > 0$ for ${\rho} > 3/7 {\lambda} = 2.47
\rho^0_{\rm   cr}.$

From (2.14b), it is found that energy density $\rho$ for phantom fluid
increases with increasing scale factor $a(t)$. It is interesting to note from
(3.5) that curvature-induced phantom fluid, obtained here, behaves effectively
as phantom violating WEC  till ${\rho} < 2.47 \rho^0_{\rm   cr}$, but phantom characterisic to violate WEC is suppressed by
cosmic tension when $\rho$ 
increases and obeys the inequality ${\rho} > 2.47 \rho^0_{\rm   cr}.$

Further, using (2.11d) in (3.3) , it is also obtained that
$$\rho + 3 P = - \frac{11\rho}{4\lambda}  - \frac{7\rho^2}{4\lambda}.  \eqno(3.6)$$

Connecting (2.28) and (3.3), it is obtained that SEC is violated when ${\rho} <
11 {\lambda}/7 = 9.07 \rho^0_{\rm   cr}  $. Also, it is found that $\rho +
3 P = 0$ for ${\rho} = 9.07 \rho^0_{\rm   cr}$ and $\rho + 3 P > 0$ for
${\rho} > 9.07 \rho^0_{\rm   cr}$.

Thus, it is obtained that (i) WEC is violated for ${\rho} < 2.47 \rho^0_{\rm
  cr}  $, (ii) for $  2.47 \rho^0_{\rm   cr} \le {\rho} < 9.07 \rho^0_{\rm   cr} $ WEC is not violated, but SEC is violated
and (iii) for ${\rho} > 9.07 \rho^0_{\rm   cr}$ neither of the two conditions is violated. Also it is
interesting to 
note that these corrections cause {\em effictive phantom divide} at
$${\rho} =  2.47 \rho^0_{\rm
  cr}. \eqno(3.7)$$  

Moreover, these results suggest that a transition from violation of SEC to
non-violation of SEC will take place at  
$${\rho} = 9.07 \rho^0_{\rm   cr} . \eqno(3.8)$$ 
Also,
universe will super-accelerate till $\rho_* < {\rho} < 2.47 \rho^0_{\rm
  cr}  ,$  accelerate when $  2.47 \rho^0_{\rm   cr} < {\rho} < 9.07
\rho^0_{\rm   cr} $ and decelerate when $9.07 \rho^0_{\rm   cr}< \rho < 11.54
\rho^0_{\rm   cr}$ as expansion of phantom phase of the universe will stop at $\rho = 11.54 \rho^0_{\rm   cr}$.

These results are also supported by (2.25b) as (2.25b) and (2.14b) yield
$$ {\ddot a} = 0.27 H_0^2 a^{5/2} \Big[0.29 -
\frac{2.68 \rho^0_{\rm cr}}{\rho} \Big]. \eqno(3.9)$$

Connecting (2.14b), (2.25a) , (2.28) and (2.29a), it is obtained that
$$ \rho = 0.73 \rho^0_{\rm cr} [0.06 + \{1.094 - 0.32 H_0 (t - 0.59 t_0) \}^2
]^{-1}. \eqno(3.10)$$

(3.7) and (3.10) yield that {\em effective phantom divide} is obtained at time
$$ t \simeq 2.42 t_0 . \eqno(3.11)$$

(3.8) and (3.10) yield that  transition time for violation of SEC to
non-violation of SEC will take place at  
$$ t \simeq 3.44 t_0 . \eqno(3.12)$$

 These results imply super-acceleration during the time
 interval $0.59 t_0 < t <  2.42 t_0$ , acceleration during the time interval
 $ 2.42 t_0 < t < 3.44 t_0$ and deceleration during the time interval $3.44
 t_0 < t < 3.87 t_0.$ Expansion, driven by phantom, will stop at time $t =
 3.87 t_0$ as $\rho_{\rm de}$ will acquire the value $2\lambda$ by this time.

When $t > 3.87 t_0$, deceleration, driven by matter, will resume and Freidmann
equation reduces to (2.17). 

\newpage
\centerline {\underline{\bf 4.  Cosmic collapse using classical approach and its quantum escape}}

\noindent \underline{\bf 4(a). Re-appearance of matter-dominance and cosmic
  collapse}

\smallskip

As mentioned above, dark energy terms are switched off, in (2.14a), at $\rho_{\rm
de} = 2 \lambda = 11.54 \rho^0_{\rm cr}$ (obtained from (2.28)), $ a = a_{\rm
pe}$ given by (2.30) when $t = t_{\rm pe} = 3.87 t_0$ given by (2.31). So, for
$t > t_{\rm pe} = 3.87 t_0$.  (2.14a) will look like 
$$ \Big(\frac{\dot a}{a} \Big)^2 =  \frac{8 \pi G}{3} \Big\{\rho^{\rm
  pe}_{\rm dm} \Big(\frac{a_{\rm pe}}{a} \Big)^3 \Big\{ 1 + {\tilde \Lambda}^{-1}
  \Big(\frac{a}{a_{\rm pe}} \Big)^3 \Big\},    \eqno(4.1a)$$
where
$$\rho^{\rm pe}_{\rm dm} = 3.68 \times 10^{-6} \rho^0_{\rm cr}, \eqno(4.1b)$$

$${\tilde \Lambda} = \frac{\rho^{\rm
  pe}_{\rm dm}}{\rho_{\Lambda}} = \frac{243 \pi G \beta \rho^{\rm pe}_{\rm dm}}{2 \alpha} = 168, \eqno(4.1c)$$
which is evaluated using (2.18d), (2.30) and (4.1b).
and
$$ \rho^{\rm pe}_{\rm  de}\Big(\frac{a}{a_{\rm pe}} \Big)^{3/4} \Big\{ 1 -
  \Big(\frac{a}{a_{\rm  pe}} \Big)^{3/4} \Big\} = 0 \eqno(4.1d)$$
as
$$ \rho^{\rm pe}_{\rm de} = 2\lambda =11.54 \rho^0_{\rm cr}. \eqno(4.1e)$$

(4.1a) integrates to
$$ a(t) = {\tilde \Lambda}^{1/3} a_{\rm pe} sinh^{2/3} \Big[ sinh^{-1} {\tilde \Lambda}^{-1/2} + 2.22
\times 10^{-4} H_0 (t - t_{\rm pe}) \Big]  \eqno(4.2a)$$
 yielding
$$ \frac{\ddot a}{a} = - \frac{2}{9} [ 2.22 \times 10^{-4} H_0 ]^2 [cosech^2
\theta (t) - 2]  \eqno(4.2b)$$
with $ \theta (t) =  sinh^{-1} {\tilde \Lambda}^{-1/2} + 2.22 \times 10^{-4} H_0 (t -
t_{\rm pe})$.  (4.2b) shows  decelerated expansion cuased  by matter
dominance till 
$$ sinh\theta (t) < 1/\sqrt{2} = 0.707 . \eqno(4.2c)$$

The result (4.2a) is obtained when expansion is driven by the term 
$$ \rho^{\rm
  pe}_{\rm dm} \Big(\frac{a_{\rm pe}}{a} \Big)^3 \Big\{ 1 + {\tilde \Lambda}^{-1}
  \Big(\frac{a}{a_{\rm pe}} \Big)^3 \Big\} $$
in FE.

Moreover, though at $a = a_{\rm pe}$,  
$$ \rho^{\rm pe}_{\rm
  de}\Big(\frac{a}{a_{\rm pe}} \Big)^{3/4} \Big\{ 1 -  \Big(\frac{a}{a_{\rm
  pe}}   \Big)^{3/4} \Big\} $$
vanishes, it will be negative for $a > a_{\rm pe}$. So for $a > a_{\rm pe}$,
  Friedmann equation (2.14a) is obtained as
$$ \Big(\frac{\dot a}{a} \Big)^2 =  \frac{8 \pi G}{3} \Big[ \Big\{\rho^{\rm
  pe}_{\rm dm} \Big(\frac{a_{\rm pe}}{a} \Big)^3 \Big\{ 1 + {\tilde \Lambda}^{-1}
  \Big(\frac{a}{a_{\rm pe}} \Big)^3 \Big\} - \rho^{\rm pe}_{\rm
  de}\Big(\frac{a}{a_{\rm pe}} \Big)^{3/4} \Big\{  \Big(\frac{a}{a_{\rm
  pe}}   \Big)^{3/4} - 1 \Big\} \Big]. \eqno(4.3)$$

Obviously, the negative terms in (4.3) will try to stop expansion and, on
sufficient growth of $a(t)$ upto $a_m$, expansion will reach it maximum such
that $ {\dot a}_{a = a_m} = 0$ and $a_m$ satisfies the equation
$$ \rho^{\rm
  pe}_{\rm dm} \Big\{\Big(\ \Big(\frac{a_{\rm pe}}{a_m} \Big)^3 + {\tilde \Lambda}^{-1}
\Big\} = 2 \lambda \Big(\frac{a_m}{a_{\rm pe}} \Big)^{3/4} \Big\{
  \Big(\frac{a_m}{a_{\rm  pe}}   \Big)^{3/4} - 1 \Big\} \eqno(4.4a)$$
being approximated as
$$ \rho^{\rm
  pe}_{\rm dm} \Lambda^{-1}
 \simeq  2 \lambda \Big(\frac{a_m}{a_{\rm pe}} \Big)^{3/4} \Big\{
  \Big(\frac{a_m}{a_{\rm  pe}} \Big)^{3/4} - 1 \Big\}. \eqno(4.4b)$$
(4.4b) yields the solution
$$ \Big(\frac{a_m}{a_{\rm  pe}} \Big) = \Big\{\frac{1}{2} \Big[1 + \sqrt{1 + 2
  \rho^{\rm   pe}_{\rm dm}/\lambda } \Big]\Big\}^{4/3} = 1 + 4.25 \times 10^{-7}. \eqno(4.4c)$$
The negative sign $(-)$ is ignored here as it yields $ a_m < a_{\rm pe},$
  which is not possible. Using (4.4c), it is obtained that
$$ \rho^m_{\rm dm} = \rho^{\rm pe}_{\rm dm} \Big[1 - 1.28 \times 10^{-6} \Big]. \eqno(4.4d)$$

Using (4.4c) in (4.2a), It is obtained that the time $t = t_m$ corresponding to
$a = a_m$ is given by
\begin{eqnarray*}
\Big[ sinh^{-1} {\tilde \Lambda}^{-1/2} &+& 2.22
\times 10^{-4} H_0 (t_m - t_{\rm pe}) \Big] \\&=& sinh^{-1} \Big[\frac{1}{4
  \sqrt{{\tilde \Lambda}}}\Big[1 + \sqrt{1 + 2  \rho^{\rm   pe}_{\rm dm}/\lambda }
\Big]^2\Big] \\&=& sinh^{-1} (0.18123) = 0.18123 . 
\end{eqnarray*}
\vspace{-1.8cm}
\begin{flushright}
(4.4e)
\end{flushright}
(4.4e) confirms deceleration during time period $ t_{\rm pe} < t < t_m$ as it
   satisfies the conditon (4.2c).

So, for $ t_{\rm pe} < t < t_m,$ (4.2a) is obtained as 
$$ a(t) = a_{\rm pe} \Big[1 + 2.22 \times 10^{-4} H_0 {\tilde \Lambda}^{1/3} a_{\rm
  pe}^{-3/2} (t - t_{\rm pe}) \Big]^{2/3} . \eqno(4.5)$$ 

Connecting (2.12b), (4.1c) and (4.4e), $t_m$ is evaluated as
$$ t_m - t_{\rm pe} = 5.32 \times 10^{-4} t_0 =  3.51\times 10^{38} {\rm
  GeV}^{-1} = 694.4 {\rm kyr}.  \eqno(4.6)$$

Further, it is interseting to note that the curve $a = a(t)$ will be
continuous at $t = t_m$, but the direction of tangent to this curve (pointed
at $a = a_m$) will
change  because it will attain its maximum at $t = t_m$ yielding ${\dot a} <
0$ for $t > t_m$, which used to be positive for $t < t_m.$ It means that
universe will retrace back at $t = t_m$ and will begin to contract. During the
contraction phase, term proportional to $a^{-3}$ will dominate over terms
proportional to $a^{3/4}$ and (4.3) will yield
$$ \frac{\dot a}{a}  \simeq - \Big[ \frac{8 \pi G}{3}\rho^{\rm
  m}_{\rm dm} \Big(\frac{a_m}{a} \Big)^3 \Big\{ 1 + {\tilde \Lambda}^{-1}
  \Big(\frac{a}{a_m} \Big)^3 \Big\}\Big]^{1/2} . \eqno(4.7)$$

On integrating (4.7), it is obtained that
$$a(t) = {\tilde \Lambda}^{1/3} a_m \{sinh[2.22 \times 10^{-4} H_0] (t_{\rm
  col} - t)\}^{2/3}, \eqno(4.8a)$$
where
\begin{eqnarray*}
t_{\rm col} &=& t_m + [2.22 \times 10^{-4} H_0]^{-1} sinh^{-1} {\tilde
  \Lambda}^{-1/2} 
  \\ &=& t_m + 3.33 \times 10^2 t_0.
\end{eqnarray*}
\vspace{-1.8cm}
\begin{flushright}
(4.8b)
\end{flushright}

(4.8a) shows that $a(t)= 0$ at $t = t_{\rm col}$ and energy density of the
   universe 
\begin{eqnarray*}
\rho &=&\Big[ \Big\{\rho^{\rm
  m}_{\rm dm} \Big(\frac{a_{\rm m}}{a} \Big)^3  + \rho^{\rm m}_{\rm
  de}\Big(\frac{a}{a_{\rm pe}} \Big)^{3/4} \Big] \\ &\simeq & \rho^{\rm
  m}_{\rm dm} \Big(\frac{a_{\rm m}}{a} \Big)^3 
\end{eqnarray*}
\vspace{-1.8cm}
\begin{flushright}
(4.9)
\end{flushright}
will diverge. It means that universe will collapse at $t = t_{\rm col}$. 

This result, suggested by the classical mechanics, is unphysical due to
  divergence of cosmic energy density. So, to have a viable physics around the epoch
  $t = t_{\rm   col}$, the diverging component of the density of the universe,
  which is 
  proportional to $ a^{-3}$, needs to be finite. As this unphysical situation
  is predicted by classical mechanics, we have no other alternative other than
  to resort to  quantum gravity. In what follows, we proceed in this direction, which is justified due to extremely high energy density and large
  curvature in the small interval of time near this particular epoch $t_{\rm   col}$. This situation is analogous to very early universe, where quantum
gravity effects are dominant. Earlier also, quantum gravity corrections were
  made in the 
equations of future universe under such circumstances to avoid finite time
singularities \cite{eil, nopl, sksgrg}.  
    
\smallskip

\noindent {\underline{\bf 4(b). Further analysis of contracting universe
  through }}

\noindent {\underline{\bf classical theory approach and conformal scalars}}

(4.9) indicates that  energy density of the contracting future universe will
      increase very high with increasing time. As a consequence, stellar
      equillibrium will break and, on sufficient rise in cosmic energy
      density,  all matter in the universe, including celestial bodies, will
      get smashed to highly-relativistic baryonic as well as non-baryonic
      elementary particles. Moreover, binding energy of celestial objects will
      be 
      released as radiation with high temperature being in the high energy
      states (though , here, universe is driven by dark energy and dark matter
      as these constitute dominant components of the cosmic energy density and
      emerge from gravitational sector on which model is based, but
      presence of non-dominant baryonic comonent is not ignored). Energy distribution of these elementary particles
      (bosons and fermions) is given by
$$ \rho_r = \rho_b + \rho_f = \frac{\pi^2}{30}\Big[g_b + \frac{7}{8} g_f\Big]
T^4   \eqno(4.10)$$  
in natural units used in this paper. Here $\rho_b(\rho_f)$ is the energy
density and $g_b(g_f)$ is the helicity for bosons (fermions). T is
the temperature. Moreover, pressure density for these particles, is obtained
as
$$ p_r = p_b + p_f = \frac{1}{3}[\rho_b + \rho_f] = \frac{1}{3}\rho . \eqno(4.11)$$  
 It shows that, on growth of energy density in the universe upto sufficiently
 high level, energy distribution will also change causing a change in the
 EOS of dominating fluid from $p = 0$ (for dark matter addressed
 in 4(a)) to EOS given by (4.11). Incorporating this change, cosmic energy
 density is obtained from conservation equation
$$ {\dot \rho_r} + 3 H (\rho_r + p_r) = 0 $$
as 
$$ \rho_r = \rho_{\rm ch} \Big(\frac{a_{\rm ch}}{a} \Big)^4, \eqno(4.12)$$    
where $\rho_{\rm ch} $ and $a_{\rm ch}$ are energy density and scale factor at
the epoch $t = t_{\rm ch}$ (when highly relativistic particles and radiation
are produced due to smash of cosmic matter at the epoch).

EOS $\rho_r = 3 p_r$ for an ideal fluid is obtained when trace of the
energy-momentum tensor components vanish. At the classicl level, this is true
for conformal scalars $\phi$ also. So, it is reasonable to think of existence of
conformal scalars $\phi$ too alongwith other fields repesenting bosons and
fermions giving energy density like (4.12). The action for conformal $\phi(x,t)$
ls given as
$$ S_{\phi} = \int {d^4x} \sqrt{-g} \frac{1}{2} \Big[ g^{\mu\nu}\partial_{\mu}\phi
\partial_{\nu}\phi - \frac{1}{6} R \phi^2 \Big] . \eqno(4.13)$$

According to above arguments, for $t \ge t_{\rm ch}$, cosmic matter will break to
elementary particles having different energy distribution (4.12), so energy
distribution (4.9) will be replaced by (4.12) in FE as  source of $\rho_{\rm
  dm} \sim a^{-3}$ will vanish. So, for $t \ge
t_{\rm ch}$, (4.7) and (4.8a) will not be valid. Now, in this stage, (4.3) is
modified as
$$ \Big(\frac{\dot a}{a}\Big)^2  \simeq  \Big[ \frac{8 \pi G}{3} \rho_{\rm ch} \Big(\frac{a_{\rm ch}}{a} \Big)^4 \Big\{ 1 + {\tilde \Lambda}_{\rm ch}^{-1}
  \Big(\frac{a}{a_{\rm ch}} \Big)^4 \Big\}\Big] . \eqno(4.14a)$$
with 
$${\tilde \Lambda}_{\rm ch} = \rho_{\rm ch}/\rho_{\Lambda} . \eqno(4.14b)$$
(4.14a) yields
$$ \frac{\dot a}{a}  \simeq - \Big[ \frac{8 \pi G}{3} \rho_{\rm ch} \Big(\frac{a_{\rm ch}}{a} \Big)^4 \Big\{ 1 + {\tilde \Lambda}^{-1}
  \Big(\frac{a}{a_{\rm ch}} \Big)^4 \Big\}\Big]^{1/2}  \eqno(4.14c)$$
replacing (4.7).

On integrating (4.14c), it is obtained that
$$a(t) = {\tilde \Lambda}^{1/2} a_{\rm ch}\Big\{sinh \Big[\sqrt{\frac{8 \pi
      G \rho_{\rm ch}}{3}}(t^{\rm ch}_{\rm col} - t) \Big]\Big\}^{1/2} , \eqno(4.15a)$$
where
$$ t^{\rm ch}_{\rm col} = t_{\rm ch} + \Big[\sqrt{\frac{3}{32 \pi
      G \rho_{\rm ch}}} \Big] sinh^{-1} {\tilde \Lambda}^{-1/2}  .\eqno(4.15b)$$

Like (4.8a), (4.12) and (4.15a) also exhibits cosmic collapse in future as $a(t) \to 0$
and $\rho \to \infty$ when $t \to t^{\rm ch}_{\rm col}$. Thus, in the changed
circumstances too, classical theory  predicts cosmic collapse in future.

All these results indicate that contracting phase of the future universe will
undergo a revival of states of the early universe, where elementary particles
in thermal equillibrium will dominate. According to the standard model of
cosmology, thermal equllibrium of these particles are maintained at and above
$10^{-3} {\rm GeV}$. So, using it in (4.10), it is obtained that
$$ \rho^{\rm ch}_r =  \frac{11 \pi^2}{60} T^4 = 1.8 \times 10^{-12} {\rm GeV}^4  \eqno(4.15c)$$

\noindent \underline{\bf 4(c). Quantum escape of collapse}

\smallskip

In what follows, a possibillity to escape cosmic collapse, obtained by
classical appproach, is probed using quantum corrections.
 
Quantum field theory suggests that  anamolous terms arise in energy-momentum
tensor components of conformal scalar $\phi$ \cite{nopl, sksgrg,
  ndb} and its trace does not vanish contrary to the results obtained through
the classical approach ( as mentioned above, trace of the energy -momentum
tensor vanishes for conformal scalars according to the classical
theory). These terms are advantageous being free from ultra-violet and
infra-red divergences with respect to energy modes. So,
quantum gravity yields anamolous terms as renormalized energy density $\rho_A$
and pressure density $p_A$ as
$$ \rho_A = \frac{N}{180 (4 \pi)^2}[ 3 H^4 + 6 H {\ddot H} + 18 H^2 {\dot H} -
3 {\dot H}^2 ] \eqno(4.16a)$$ 
and
$$ p_A = - \frac{N}{360 (4 \pi)^2}(6 H^4 + 8 H^2 {\dot H}) - \frac{N}{180 (4
  \pi)^2} [2 {\dddot H} + 12 H {\dot H} + 18 H^2 {\dot H} + 9 {\dot H}^2 ],
  \eqno(4.16b)$$  
with  finite positive integers $N$ being the number of conformal scalars.

On making quantum corrections, (4.14a) is obtained as
$$\frac{3}{8 \pi G} H^2 = 
  (\rho_r + \rho_A) + \rho^q_{ \Lambda} \eqno(4.17)$$
using (4.14b) and   $\rho^q_{ \Lambda}$ as $\rho_{\Lambda}$ after quantum
  corrections.

Due to correction terms in (4.17), $a(t)$ (being the solution of this
equation) will be different from (4.8a) or (4.15a). But, the required solution
is not possible unless $\rho_A$ is obtained as a suitable function of
$a(t)$. Under these circumstances, there is an alternative approach to assume
a solution  and find conditions to make this soluition capable to
yield physically viable results and satisfy (4.17). (4.8a) and (4.15a)
show  that  diveregence is caused by dominating component of cosmic energy density when $ a \to 0$ as $t \to t^{\rm ch}_{\rm col}$. Now, if quantum corrections
remove this divergence, $a(t)$ should be finite at $t = t^{\rm ch}_{\rm col}$. In
cosmology, mostly $a(t)$ is found either as a power-law solution or as an
exponential solution of FE. According to investigations here, $t = t^{|rm ch}_{\rm
  col}$  is the epoch where this universe is expected to end. So, if $a(t)
\sim ( t^{\rm ch}_{\rm col} - t)^n $ for the contracting universe for $t > t_{\rm ch}$,
diveregence can not be removed. It can be checked from (4.16a) that $\rho_A$
itself will diverge for this form of $a(t)$. So, these arguments suggest to choose an exponential form of
$a(t)$ during the quantum regime as it is possible for this form to yield
finite value $ a_{\rm col} = a(t^{\rm ch}_{\rm col})$.

With this motivation, for $t > t_c$, solution of (4.17) is taken
as 
$$ a^q(t) = a_{\rm col} exp [|\{D (t^{\rm ch}_{\rm col} - t)\}| + C_1 \{D (t^{\rm ch}_{\rm
  col} - t)\}^{\gamma}]  \eqno(4.18)$$
with $\gamma$ being a positive real number, $C$ being an arbitrary
  dimensionless constant
  and $D$ being a constant of mass dimension used for dimensional
  correction. Here $t = t_c$ is the epoch
  near $t = t^{|rm ch}_{\rm col}$, when quantum gravity effects  begin to dominate. In
  what follows, it is shown that (4.18) is physically valid satisfying (4.17)
  and numerical values $C_1, D , \gamma$ and $a_{\rm col}$ are determined.

From (4.18), it is obtained that
\begin{eqnarray*}
 H &=& - D - {\gamma} C_1 D [D (t^{\rm ch}_{\rm col} - t)]^{({\gamma} -1)} , \\
 {\dot H} &=&  {\gamma}({\gamma} -1) C_1 D^2 [D (t^{\rm ch}_{\rm col} - t)]^{({\gamma}
    -2)} , \\
 {\ddot H} &=& - {\gamma}({\gamma} -1)({\gamma} -2) C_1 D^3 [D (t^{\rm ch}_{\rm col} -
t)]^{({\gamma} -3)}, \\
 {\dddot H} &=&  {\gamma}({\gamma} -1)({\gamma} -2)({\gamma} - 3)C_1 D^4 [D
(t^{\rm ch}_{\rm col} - t)]^{({\gamma} - 4 )}.  
\end{eqnarray*}
\vspace{-3.4cm}
\begin{flushright}
(4.19a,b,c,d)
\end{flushright}
\vspace{1.5cm}

From (4.16a) and (4.16b), we find that terms containing $H$ and ${\dot H}$ are
dominant in $\rho_A$ and $p_A$. So,  minimum possible value for $\gamma$ is  taken as
$$ \gamma = 2, \eqno(4.20)$$
because $H $ and ${\dot H}$ do not vanish even at $t = t^{\rm ch}_{\rm col}$ for this value
of $\gamma$. But 
$${\ddot H} = 0 = {\dddot H}. \eqno(4.21 a,b)$$ 
Also, at $t = t^{\rm ch}_{\rm col}$
$$ H = - D   \quad {\rm and }\quad  {\dot H} =  2 C_1 D^2, \eqno(4.22 a,b)$$ 
which is evaluated using
$$\lim_{\epsilon \to 0} [D (t^{\rm ch}_{\rm col} - [t^{\rm ch}_{\rm col} -
\epsilon])]^{({\gamma} -2)} = \lim_{\epsilon \to 0} [D (t^{\rm ch}_{\rm col} - [t^{\rm ch}_{\rm
  col} + \epsilon])]^{({\gamma} -2)} = 1 \eqno(4.23)$$
for $\gamma = 2$.

Now, it is important to check whether $a^q(t)$, given by (4.18), satisfies
(4.17) with $\rho_A$ from (4.16a). Connecting (4.16a), (4.17), (4.18)
and (4.19a)-(4.19d), it is obtained that
\begin{eqnarray*}
&&\frac{3}{8 \pi}M_P^2 D^2 [1 + 4 C_1 \{D (t^{\rm ch}_{\rm col} - t)\} + 4 C_1^2 \{D (t^{\rm ch}_{\rm
  col} - t)\}^2 ] = \rho_{\rm dm} [1 - \\&& 4 |\{D (t^{\rm ch}_{\rm col} - t)\}| + 4 (2 +  C_1) \{D (t^{\rm ch}_{\rm col} - t)\}^2 ] + \rho_A + \frac{N D^4}{60 (4\pi)^2} [(1 - 4 C_1^2)\\&&
+ 8 C_1 \{D (t^{\rm ch}_{\rm col}
- t)\} + 24 C_1^2 \{D (t^{\rm ch}_{\rm col} - t)\}^2 ] 
\end{eqnarray*}
\vspace{-1.8cm}
\begin{flushright}
(4.24)
\end{flushright}
approximating upto second order only as $|D (t^{\rm ch}_{\rm col} - t)| < 1$ ,
because, according to above arguments, quantum gravity is effective near the
epoch $t = t^{\rm ch}_{\rm col}$.
Comparing constant terms and coefficients of $\{D (t^{\rm ch}_{\rm col} - t)\}$ as
well as $\{D (t^{\rm ch}_{\rm col} - t)\}^2$ in (4.24), it is obtained that
\begin{eqnarray*}
\frac{3}{8 \pi}M_P^2 D^2 &=& \rho_r|_{t = t^{\rm ch}_{\rm col}} + \rho_{\Lambda} + \frac{N D^4}{60 (4\pi)^2}
(1 - 4 C_1^2), \\ \frac{3}{2 \pi} C_1 M_P^2 D^2 &=& - 4 \rho_r|_{t = t^{\rm ch}_{\rm col}} + \frac{8 N
  D^4}{60 (4\pi)^2} C_1, \\ \frac{3}{2 \pi} C_1^2 M_P^2 D^2 &=& 
4 (2 +  C_1) \rho_r|_{t = t^{\rm ch}_{\rm col}} + \frac{24 N
  M_P^4}{60 (4\pi)^2} C_1^2
\end{eqnarray*}
\vspace{- 3.2cm}
\begin{flushright}
(4.25a,b,c)
\end{flushright}
\vspace{2cm}

Incorporating (4.16a) and (4.25a,b,c) in (4.18), it is obtained that
\begin{eqnarray*}
\rho_r + \rho^q_{ \Lambda} &=& \frac{3 }{8 \pi G}H^2 - \rho_A \\ &=&
\frac{3 }{8 \pi } M_P^2 D^2 + \frac{N D^4}{60 (4\pi)^2}
(1 - 4 C_1^2) .
\end{eqnarray*}
\vspace{-1.8cm}
\begin{flushright}
(4.25d)
\end{flushright}
 yielding $(\rho_r + \rho^q_{ \Lambda})$ finite at $t = t^{\rm ch}_{\rm col}$.

Planck scale is the fundamental scale. It suggests the largest energy mass
scale as Planck mass $M_P$. At this  scale, energy density is obtained $\sim
10^{75} {\rm   GeV}^4$. As, at $t = t^{\rm ch}_{\rm col}$,  finite $\rho_r$
is obtained through quantum corrections.  As, at $t = t^{\rm ch}_{\rm col}$,
we obtain finite $\rho_r$ due to quantum corrections, so 
$$ \rho_r|_{t = t^{\rm ch}_{\rm col}} = \frac{1}{r} M_P^4 \eqno(4.26)$$
where $r > 1$ is the real number.

(4.25b) yields
$$ \frac{C_1 }{2 \pi} M_P^2 D^2 \Big[ 3 - \frac{N}{60 \pi} \Big] = - 4 \rho_r|_{t = t^{\rm ch}_{\rm col}}. \eqno(4.27)$$

From (4.26) and (4.27), it is obtained that
$$ \frac{1 }{2 \pi} \Big[ 3 - \frac{N}{60 \pi} \Big] C_1 = -
\frac{4}{r}. \eqno(4.28a)$$
and
$$ D = M_P \eqno(4.28b)$$
as $D$ has mass dimension and $C$ is dimensionless.

Using (4.26), (4.28b) in (4.25c), we obtain
$$ \frac{3 }{2 \pi}\Big[\frac{N}{60 \pi} - 1\Big]C_1^2 + \frac{4}{r} C_1 +
\frac{8}{r} = 0. \eqno(4.29)$$

(4.29) yields the real value of $C$, when $ N < 60 \pi$. So, we can take 
$$ N = 188 . \eqno(4.30)$$

Connecting (4.28a) and (4.30), $C$ is obtained as
$$ C_1 = - \frac{3.99 \pi}{r} . \eqno(4.31)$$

(4.29) and (4.31) yield 
$$ r = 2 \pi. \eqno(4.32)$$
and
$$ C_1 = - 1.995. \eqno(4.33)$$ 

Connecting (4.25a), (4.26), (4.28b), (4.30), (4.32) and (4.33), we obtain
$$ \rho^q_{\Lambda} = 2.56 \times 10^{75} {\rm GeV}^4 . \eqno(4.34)$$

Thus, it is found that vacuum energy density $\rho^q_{\Lambda}$, given by
 (4.34), is extremely high and rises by $10^{129}$ compared to
$\rho_{\Lambda}$ given by (2.18e) on making quantum correction. It means that
quantum gravity feeds cosmological constant so well that it gets extremely fattened. Thus, it is
interesting to see that, during quantum regime, cosmological constant is
extremely heavy, but it is almost negligible in classical domain.

As $a^q(t)$, given by (4.18), is valid during quantum dominance and it will have
its maximum at $t = t_c$, if quantum corrections begin to dominate at this epoch. So, from (4.19a), (4.20) (4.28b) and (4.33), we obtain
$$ H = - M_P + 3.99  M_P [M_P (t^{\rm ch}_{\rm col} - t_c)] = 0 \eqno(4.35)$$
yielding 
$$t_c = t_{\rm col} - 0.25 t_P \eqno(4.36)$$
with $ t_P = M_P^{-1}$ being the Planck time. Thus, quantum gravity
corrections will be effective for
$$ |M_P (t_{\rm col} - t_c) | < 0.25 . \eqno(4.37)$$

(4.18), (4.20), (4.34b) and (4.36) yield
$$ a_c = 1.13 a_{\rm col} . \eqno(4.38)$$

(2.30),(4.1b), (4.4d) and (4.15a) yield
$$a_{\rm ch}  = 1.468 \times 10^{-12}. \eqno(4.39)$$ 
From (4.1c), (4.15a), (4.15c),(4.36) and (4.39) , we obtain
$$ a_c = 1.875 \times 10^{-34}. \eqno(4.40)$$
Further, (4.38) and (4.40) imply
$$ a_{\rm col} = 1.659 \times 10^{-34}. \eqno(4.41)$$

Thus, solution of FE (4.17) with quabtum currection is obtained as 
$$ a^q(t) = 1.659 \times 10^{- 34}  exp [|\{M_P (t^{\rm ch}_{\rm col} - t)\}| - 1.995 \{M_P
  (t^{\rm ch}_{\rm   col} - t)\}^2]  \eqno(4.42)$$

(4.41) yields $H < 0$ for $t < t_{\rm col}$ and $H = M_P [1 - 3.99 \{M_P(t -
       t^{\rm ch}_{\rm col} \}] >0 $ for $t > t_{\rm
       col}$. It shows contracting universe for $0.25 t_P \le t < t^{\rm ch}_{\rm col}$
       and expanding universe for $ t^{\rm ch}_{\rm col} < t \le 0.26 t_P$ during
       quantum regime.
It means that future universe will pass through the narrow quantum bridge from
       pre-collapse to post-collapse siate avoiding cosmic
       collapse suggested by classical mechanics.

\bigskip
\centerline{\underline{\bf 5. Summary }}

\smallskip

Here, cosmolgy of the late and future universe is obtained from $f(R)-$
gravity, obtained by adding higher-order curvature terms $R^2$ and $R^3$ to
the Einstein-Hilbert linear in scalar curvature $R$. It is explained in the
first section on introduction that $f(R)-$ gravity cosmology is different from
$f(R)-$ dark energy models \cite{snj}. Moreover, problems of $f(R)-$ dark
energy models, pointed out in \cite{lds}, do not appear in $f(R)-$ gravity
cosmology \cite{sks07a, sks06, sks07b, sks08a, sks08b} as well as in the
present model. Here, curvature scalar contributes to both geometrical and
physical components of the theory. Thus, it plays dual role as a geometrical
as well as physical fields, which was obtained earlier in \cite{sksr}.

Here, it is found that, in the late universe, for the red-shift $z < 1089$,
dark matter term emerge spontaneously and phantom dark energy emerge as
imprints of linear and non-linear terms of curvature. It is found that, during
$0.36 < z < 1089$, dark matter dominated and universe expanded with
deceleration as $t^{2/3}$. A transition from deceleration to acceleration took
place at $z = 0.36$ and at time $t = 0.59 t_0$ ($t_0$ being the present age of
the universe). This transition is caused by dominance of curvature-induced
phantom dark energy over curvature-induced dark matter. Dark energy gives
anti-gravity effect and phantom dark energy exhibits this effect more strongly
due to violation of WEC. So, phantom energy puts high jerk causing
super-acceleration. 

$f(R)-$ gravity inspired Friedmann equation, obtained here, contains two terms
(i) $8 \pi G \rho_{\rm dm}/3$ ( with $\rho_{\rm dm}$ being the dark matter
density) and (i) $(8 \pi G \rho_{\rm de}/3) [1 - \rho_{\rm de}/2 \lambda] $ (
with $\rho_{\rm de}$ being the phantom dark energy density). Here $\lambda =
5.777 \rho^0_{\rm cr}$ is called the cosmic tension. It is interesting
to see 
that Friedmann equation, obtained here, contains a correction term  $ - 4 \pi
G \rho^2_{\rm de}/3\lambda$  analogous to such term in Friedmann equations from
RS-II model of brane-gravity \cite{rm} and loop quantum cosmolgy\cite{ms}. This
correction is not effective in the present universe as $ \rho^0_{\rm de} <<
2\lambda$ as well as it is ineffective till $ \rho_{\rm de} << 2\lambda$. But,
as $ \rho_{\rm de}$ will increase in future with the growing scale factor
$a(t)$, effect of this term will increase. It is found that, on sufficient
growth of $ \rho_{\rm de}$, the {\em effective EOS} does not violate WEC (which
characterizes phantom), but violates SEC. On further increase in $ \rho_{\rm
  de}$, even SEC is not violated. Thus, in the beginning of the
phantom phase, universe will super-accelerate during the period $0.59 t_0 < t
< 2.42 t_0$, it will accelerate during the period $2.42 t_0 < t < 3.44 t_0$
and, during the period $3.44 t_0 < t < 3.87 t_0$, universe will decelerate even
during the phantom phase. Phantom-dominance will end when $\rho = 2 \lambda =
11.54 \rho^0_{\rm cr}$ at time $t = 3.87 t_0$. As a consequence, re-dominance
of dark matter will begin giving decelerated expansion. But, as universe will
expand, growth of $\rho_{\rm de} \sim a^{3/4}$ will also continue giving
$\rho_{\rm de} > 2 \lambda$. It causes the term  $(8 \pi G \rho_{\rm de}/3) [1
- \rho_{\rm de}/2 \lambda] $ to switch over from positive to negative. Growth of this negative term  will try to slow down expansion more rapidly. As a
result, universe will reach a state, when expansion will stop and scale factor
will acquire its maximum value in finite time $ t_m = 3.87 t_0 + 694.4 {\rm
  kyr}.$ When $t > t_m$, universe will retrace back and contract. Results,
obtained here, show that contraction will continue for sufficiently long
period $333 t_0 $ and will collapse at time $t_{\rm col} = 336.87 t_0 +  694.4 {\rm
  kyr},$ where energy density of the universe will diverge and scale factor
will vanish. These results are obtained using prescriptions of classical
mechanics. In this paper, it is probed further whether quantum gravity
corrections can save the universe from the menace of collapse.

It is argued, here, that as in the very small time priod close to $t = t_{\rm
  col}$, energy density is extremely high and curvature is very large, quantum
  gravity effects will be dominant. It is analogous to the state of very early
  universe, where energy density is extremely high and curvature is very
  large. So, quantum gravity has a crucial role. Motivated by these
  arguments, quantum corrections are used in the future universe near time of
   collapse and it is found here that expected collapse of the future universe
  can be avoided. Moreover, results show that contracting universe, in
  pre-collapse era  will  expand with super-acceleration in the post-collapse
  era. 

Interestingly, energy density due to cosmological constant $\rho^q_{\Lambda}$,
during quantum regime, is found $10^{129}$ times $\rho_{\Lambda}$ during
  regime of classical mechanics. It means that quantum gravity nourishes the
  cosmological constant well and makes it very strong, otherwise it is
  very weak under the rule of classical mechanics. So, when $\rho_{\Lambda}$
  is out of quantum domain, it falls down immediately by $129$ order.
 
The state of the universe near the epoch $t = t_{\rm col}$ can be
realized as revival of the state of early universe in future due to high
energy density, large curvature and dominance of quantum gravity effects. If
it is true, study of this state of the future universe can unveil many hidden
knowledge of the early universe. For example, here, results of the
cosmological constant near collapse time explains largeness of $\rho_{\Lambda}$
in the very early universe and its extremely small value in the present
universe. It is important to see that this result is obtained without any
fine-tuning.


\begin{thebibliography}{14}

\smallskip

\bibitem{sp}
 S. J. Perlmutter $et$ $al.$, Astrophys. J. {\bf 517},(1999)565;
 astro-ph/9812133;  D. N. Spergel $et$ $al$,  Astrophys J. Suppl. {\bf 148}
 (2003)175[ astro-ph/0302209] and references therein.

\smallskip
\bibitem{ag}
 A. G. Riess $et$ $al$, Astrophys. J. {\bf 607}, (2004) 665 [
 astro-ph/0402512].

\smallskip

\bibitem{ejc}
 E.J.Copeland, M.Sami and S. Tsujikawa, Int. J. Mod. Phys. D, {\bf
 15},(2006)1753 [hep-th/0603057] and references therein. 


\smallskip

\bibitem{sksch}
S.K.Srivastava, Phys.Lett. {\bf B 619} (2006)1 [astro-ph/0407048]
\smallskip

\bibitem{cap}
S. Capozziello, V.F.Cardone, S.Carloni and A.Troisi, Int. J. Mod. Phys. D, {\bf
 12},(2003)1969; S.M. Carroll, V.Duvvuri, M. Trodden and M.S.Turner,
Phys. Rev.D, {\bf  70},(2004) 043528.

\smallskip
\bibitem{snj}
S. Nojiri and S.D.Odintsov, Int.J. Geom. Meth. Mod. Phys. {\bf 4},(2007)115
[hep-th/0601213 ]and references therein.

\smallskip

\bibitem{lds}
L. Amendola, D. Polarski and S. Tsujikawa, Phys.Rev. Lett. {\bf 98} (2007)
131302 [astro-ph/0603703] ; L. Amendola, D. Polarski, R.Gannouji and S. Tsujikawa,  Phys.Rev.D,
{\bf 75} (2007) 083504 [gr-qc/0612180].

\smallskip
\bibitem{sks07a} 
S.K.Srivastava, Int. J. Mod. Phys. A {\bf 22} (6) (2007) 1123-34
[hep-th/0605010] .

\smallskip
\bibitem{sks06}
S.K.Srivastava, Phys.Lett.B,{\bf 643} (2006) p.1-4 [astro-ph/0608241].

\smallskip
\bibitem{sks07b}
S.K.Srivastava, Phys. Lett. B {\bf 648}
  (2007) 119-126 [astro-ph/0603601]..

\smallskip
\bibitem{sks08a}
S.K.Srivastava, to appear in Int. J. Mod. Phys. D [ astro-ph/0602116].

\smallskip
\bibitem{sks08b}
S.K.Srivastava, to appear in Int. J. Theo. Phys.[arXiv:0706.0410 [hep-th]].



\smallskip
\bibitem{rm}
 R. Maartens, gr-qc/0312059.


\smallskip
\bibitem{ms}
M. Sami, P. Singh and S. Tsujikawa, Phys. Rev.D, {\bf 74} (2006)043514[gr-qc/0605113].



\smallskip
\bibitem{eil}
E. Elizalde, S. Nojiri and  S.D. Odintsov, Phys. Rev. D {\bf 70}(2004)
043539. 

\smallskip

\bibitem{nopl}
Nojiri, S., Odintsov, S.D.,(2004), Phys. Lett. B, {\bf 595}, 1 (2004)
hep-th/0408170].

\smallskip
\bibitem{sksgrg}
Srivastava, S.K., (2007), Gen. Rel. Grav., {\bf 39}, 241.


\smallskip
\bibitem{ad}
A.D. Miller $et$ $al$ , Astrophys. J. Lett. {\bf 524} (1999) L1; P. de
Bernadis $et$ $al$ , Nature (London){\bf 400} (2000) 955; A.E. Lange $et$ $al$
, Phys. Rev.D{\bf 63} (2001) 042001; A. Melchiorri $et$ $al$ ,
Astrophys. J. Lett. {\bf 536} (2000) L63; S. Hanay $et$ $al$ , 
Astrophys. J. Lett. {\bf 545} (2000) L5.


\smallskip
\bibitem{abl} 
A.B. Lahnas, N.E. Mavromatos and D.V. Nanopoulos,  Int. J. Mod. Phys. D, {\bf 
  12(9)}, 1529 (2003).  

\smallskip
\bibitem{ndb}
N.D.Birrel and P.C.W.Davies, ``Quantum Fields in Curved Spaces'', Cambridge
University Press, Cambridge (1982).


\bibitem{sksr}
 S.K.Srivastava and K.P.Sinha; Phys.Lett.B, {\bf 307} (1993) 40;
  Pramana, {\bf 44} (1993) 333;  Jour. Ind. Math. Soc.{\bf 61},80 (1994);
  Int.J.Theo.Phys., {\bf 35} (1996) 135;  Mod.Phys.Lett.A, {\bf 12} (1997)
  2933; S.K.Srivastava; Il Nuovo  Cimento B, {\bf 113} (1998) 1239;
  Int.J.Mod.Phys.A, {\bf 14} (1999) 875; Mod.Phys.Lett.A, {\bf 14} (1999)
  1021; Int.J.Mod.Phys.A, {\bf 15} (2000) 2917; Pramana, {\bf 60} (2003) 29;
 hep-th/0404170; gr-qc/0510086. 
\end{thebibliography}
\end{document}